# Comparison of OpenMP & OpenCL Parallel Processing Technologies

Krishnahari Thouti

Department of Computer Science & Engg.
Visvesvaraya National Institute of Technology,
Nagpur – 440010 (India)

S.R.Sathe

Department of Computer Science & Engg.
Visvesvaraya National Institute of Technology,
Nagpur – 440010 (India)

*Abstract*—This paper presents a comparison of OpenMP and OpenCL based on the parallel implementation of algorithms from various fields of computer applications. The focus of our study is on the performance of benchmark comparing OpenMP and OpenCL. We observed that OpenCL programming model is a good option for mapping threads on different processing cores. Balancing all available cores and allocating sufficient amount of work among all computing units, can lead to improved performance. In our simulation, we used Fedora operating system; a system with Intel Xeon Dual core processor having thread count 24 coupled with NVIDIA Quadro FX 3800 as graphical processing unit.

## I. INTRODUCTION

Nowadays, Quad-core, multi-core & GPUs [1] have already become the standard for both workstations and high performance computers. These systems use aggressive multi-threading so that whenever a thread is stalled, waiting for data, the thread can efficiently switch to execute another thread. Achieving good performance on these modern systems requires explicit structuring of the applications to exploit parallelism and data locality.

Multi-core technology offers very good performance and power efficiency and OpenMP [2] has been designed as a programming model for taking advantage of multi-core architecture. The problem with GPU is that, their architecture is quite different to that of a conventional computer and code must be (re)written to explicitly expose algorithmic parallelism. A variety of GPU programming models have been proposed in [3 - 5].

The most popular development tool for scientific GPU computing has proved to be CUDA (Compute Unified Device Architecture) [6], provided by the manufacturer NVIDIA for its GPU products. However, CUDA is not designed for heterogeneous systems, while OpenCL programming model, by the Kronos Group [7] supports cross-platform, parallel programming of heterogeneous processing systems. The architectural details of multi-core and GPUs are explained in next section.

Given, a diversity of high-performance architectures, there is a question of which is the best fit for a given workload and extent to which an application benefit from these systems, depends on availability of cores and other workload parameters. This paper addresses these issues by implementing parallel algorithms for the four test cases and compares their performance in terms of time taken to execute and percentage of speed-up factor achieved.

In Section II, we present parallel computing paradigm. We then present architectural framework for Multi-core and GPU architectures in Section III & IV.

Experimental results are presented in Section V. Section VI presents related work done and conclusion and future scope are discussed in Section VII.

## II. PARALLEL COMPUTING PARADIGM

Parallel computing [8] depends on how the processors are connected to memory. The way of system connection can be classified into shared or distributed memory systems, each of these two types are discussed as follows:-

### A. Shared Memory System

In such a system, a single address space exists, within it every memory location is given a unique address and the data stored in memory are accessible to all processing cores. The processor $P_i$ reads the data written by processor $P_j$. Therefore, in order to enforce consistency, it is necessary to use synchronization.

The OpenMP is one of the popular programming languages for the shared memory systems. It provides a portable, scalable and efficient approach to run parallel programs in C/C++ and FORTRAN.

In OpenMP, a sequential programming language can be parallelized with pre-processor compiler directive #pragma omp in C and $OMP in FORTRAN.

### B. Distributed Memory System

In such a system, each processor has its own memory and can only access its local memory. The processors are connected with other processors via high-speed communication links. MPI (Message Passing Interface) [9] provides a practical, portable, efficient and flexible standard for message passing across distributed memory systems.

We limit our discussion to shared memory systems. Based on above classification, we classify systems as Multi-core systems and Many-core systems or GPGPU devices [1].

## III. ARCHITECTURAL FRAMEWORK – MULTICORE & OPENMP PROGRAMMING MODEL

The present typical multi-core architecture is shown in Figure 1. It consists of a number of processing cores, each having a private level one (L1) data and instruction cache, L2 cache, which are attached via a bus interconnect to shared level three (L3) cache.

Each core supports multi-threading, which allows sharing of several micro-architectural resources between threads e.g. L1 caches, physical registers, and execution units.

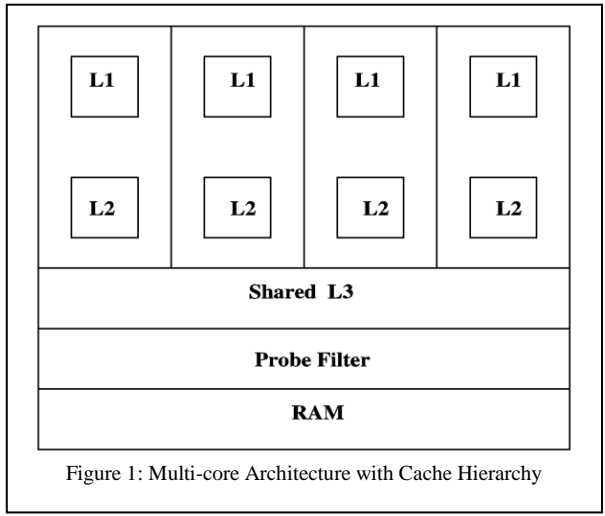

Figure 1: Multi-core Architecture with Cache Hierarchy

Benefits of multi-core systems can be obtained by Open MP Programming model. OpenMP is a specification of compiler directives, library routines, and environmental variables that provides an easy parallel programming model portable across shared memory architecture.

OpenMP is a set of compiler directives (# pragma) and callable runtime library routines that express shared memory parallelism [10]. The directive itself consists of a directive name and followed by clauses. OpenMP programs execute serially until they encounter the "parallel" directive. This directive is responsible for creating group of threads. The exact number of threads can be specified in the directive, set using an environmental variable, or at run-time using OpenMP functions. The main thread that encounters the "parallel" directive becomes the "master" of this group of threads and is assigned the thread id 0. There is an implicit barrier at the end of parallel region. The master thread with thread id 0 collects results from other threads and executes serially from that point on.

## IV. ARCHITECTURAL FRAMEWORK – GPU & OPENCL PROGRAMMING MODEL

GPUs i.e. Graphic Processing units are the basic building blocks for high performance computing but its programming complexity pose a significant challenge for developers. To improve the programmability of GPUs, the Open CL (Open Computing Language) [7] has been introduced. Open CL is an industry standard for writing parallel programs to execute on the heterogeneous platforms like GPU devices. OpenCL (Open Computing Language) is a low-level API for heterogeneous computing that runs on CUDA architecture.

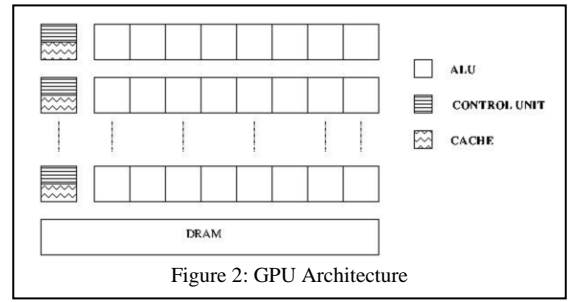

Figure 2: GPU Architecture

NVidia GPUs comprises of array of multithreaded Streaming Multiprocessors (SMs) and each one consists of multiple Scalar Processor (SP) cores, a multithreaded instruction unit, and on-chip shared memory. The SMs creates, manages, and execute concurrent threads in hardware with zero scheduling overhead.

In short, we say, following are the steps to initialize an OpenCL Application.

Set Up Environment – Declare OpenCL context, choose device type and create the context and a command queue.

Declare Buffers & Move Data – Declare buffers on the device and enqueue input data to the device.

Runtime Kernel Compilation – Compile the program from the kernel array, build the program, and define the kernel.

Run the Program – Set kernel arguments and the work-group size and then enqueue kernel onto the command queue to execute on the device.

Get Results to Host – After the program has run, read back result array from device buffer to host memory.

## V. EXPERIMENTAL RESULTS

In this section, we present experimental results. For experimental setup, we have tested our system on four test cases. We compare the performance of these test cases with the OpenCL code on the GPU and on a multi-core CPU with Open MP support.

The host machine used has Intel Xeon 2.67GHz Dual processors with 12Mb L3 cache. Each processor has Hyper-Threading [11] technology such that, each processor can execute simultaneously instructions from two threads. Overall numbers of cores are 12 and because of hyper threading thread count of host machine equal 24.

Each core of two processors has 32KB L1 Data cache, 256 KB L2 cache shared between 2 threads of that core. In addition to that, there is 12MB L3 cache shred among all the threads.

The GPU device used in our experiment was NVidia Quadro FX3800. The device has 192 processing cores with 1 GB 256 bit memory interface and memory bandwidth of 51.2 GB/sec. The GPU device was connected to CPU through X58 I/O Hub PCI Express. The environment used was Fedora-x86_64 and kernel version is - 2.6. Gcc version is 4.6.

OpenCL-1.0 was used for compiling OpenCL programs by providing "-lOpenCL" as compile time option for gcc compiler.

We have used "gettimeofday ()" library routine to measure time taken to execute test problems. "start" and "end" time recorded and execution time is calculated as shown below in Program Listing-1:

```
gettimeofday (&start, NULL );
    Backtrack (0, 0, 0, 0);
    gettimeofday (&end, NULL );
    time = (end.tv_sec-start.tv_sec) + (float) (end.tv_usec - start.tv_usec) * 0.000001;
    printf ("Time = %f\n", time );
```

Program Listing 1: Measuring CPU Time

### A. Matrix Multiplication

We consider the problem of computing the product C = A*B of two large, dense, matrices. A straight forward matrix multiplication performs scalar operations on data items. We choose matrix multiplication, because of following two reasons:

- Matrix Multiplication is widely used in Image processing applications

- Matrix Multiplication is a fundamental parallel algorithm with respect to data locality, cache coherency etc.

| Matrix Order | Sequential | OpenMp | OpenCL |
|---|---|---|---|
| 1024 | 6.04 | 0.71 | 1.64 |
| 2048 | 136.14 | 18.39 | 2.05 |
| 3072 | 345.06 | 40.84 | 2.45 |
| 4096 | 1261.29 | 177.79 | 3.66 |
| 5120 | 2819.54 | 328.52 | 5.53 |
| 6144 | 5023.87 | 593.13 | 8.4 |

Table 1: Execution Time for Matix Multiplication of Sequential, OpenMp, OpenCL Version

| Matrix Order | Seq-to-OpenMp | Seq-to-OpenCL |
|---|---|---|
| 1024 | 8.51 | 3.68 |
| 2048 | 7.40 | 66.41 |
| 3072 | 8.45 | 140.85 |
| 4096 | 7.09 | 344.62 |
| 5120 | 8.58 | 509.86 |
| 6144 | 8.47 | 598.08 |

Table 2: Speed-Up of Sequential to openMP & OpenCL

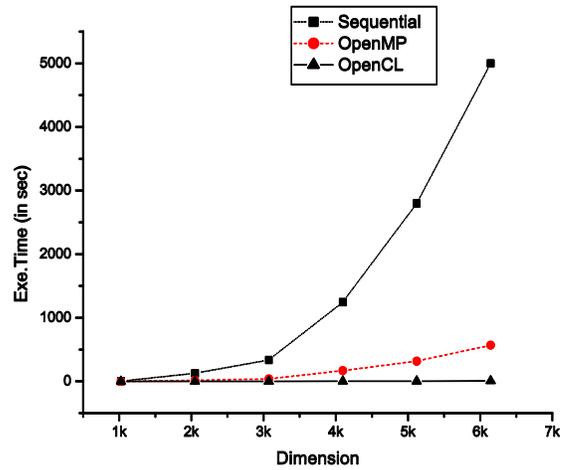

Figure 3: Execution Time v/s Dimension of Matrix

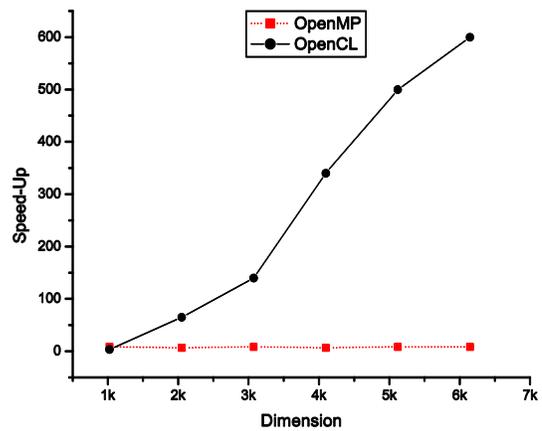

Figure 4: Speed-Up Comparison

As the dimensions of the matrix increase, the execution time for sequential algorithm also increases by manifold as shown on Figure 3. After analyzing Table -1 and 2, Figure 3 and 4, we conclude that, given the Multi-core architecture, OpenMP shows good improvements for smaller matrix dimensions but as the matrix dimension increases OpenCL gives very good Speed-Up factor and very less execution time. This can be evident from Table 1 & 2 that for matrix dimension 1024 OpenMP is much better than OpenCL. But for matrix dimension above 1024, OpenCL gives very good performance. Note that values shown in Table 1 & 2 are obtained after performing the experiment for nearly 10 – 20 runs.

### B. N-Queens Problem

The n-queen problem is a classic problem of placing n-chess queens on chessboard so that no two queens attack each other.

The most obvious way to solve this problem consists of trying systematically all ways of a placing N-Queens on a chessboard, checking each time to see whether a solution has been obtained. But this approach will take very large time to arrive at solution. Backtracking is an approach to solve this problem. But backtracking takes exponential time-complexity. Because of this reason, it is very interesting to parallelize this problem.

Recursion is a peculiar property of backtracking. The earlier version of OpenCL doesn't support recursion. Support for recursion is introduced in OpenMP 3.0 specifications by "task "clause. However, we find that there is no significant improvement in performance, since most of the code to be parallelized is kept in critical section region as shown below:

```
int put(int Queens[], int row, int column)
{
    Queens[row]=column;
    if(row==N-1)
    pragma omp critical
    {    solutions++;   }
    }
    else{
    for(i=0; i<N; i++){ put(Queens,row+1,i);}
    }
    return solutions;
}
```

**Program Listing 2: OpenMP code for N-Queens**

Therefore, we have taken into consideration of sequential v/s OpenCL code to evaluate N-Queen problem. For larger values of N (> 23), the number of solutions and time to solve the given problem is still not known [12, 13].

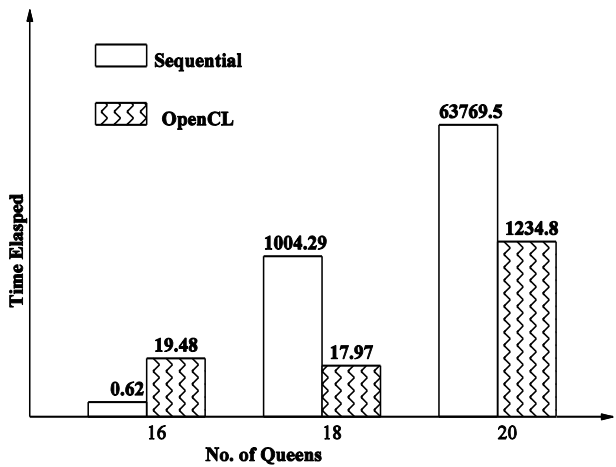

Figure 5: Performance of N-Queen Problem

As shown in Figure 5, we have taken only practical cases where solutions are available within stipulated time. For N=18, sequential took nearly 16.75Minutes (1004.29sec) whereas OpenCL took only 17.97 seconds to generate all correct results. The power of OpenCL can be observed in cases for N >=20. For N=20, sequential program took nearly 17.72Hrs (63769.5sec) whereas OpenCL took only 20.58Min (1234.88sec). Speed-up achieved is enormous.

The graph shown in Figure 5 is not up to the scale.

### C. Image Convolution

The convolution of images is a commonly used technique for image filtering. It is best described as a combining process that copies one image into another. Any number of filters may be applied to an image by convolving the filter mask with the original image. The equation for image convolution is given by

$$Out(i, j) = \sum_{m=0}^{M-1} \sum_{n=0}^{N-1} In(m,n)Mask(i-m, j-n)$$

Where In is the input image, Mask is the convolution mask, and Out is the output image. The dimension of the image is M x N, the Mask image is smaller than image size; may be padded with zeroes to allow for consistency in indexing.

The convolution technique consists of the following steps:

- Select a pixel in original image to convolute
- Apply mask to the pixel by reading the selected pixel's neighbor
- Write the new values to the out image

The convolution algorithm will generate results that are greater than the range of original values of the input image. For this, scaling operation is performed to restore the result to same gray level range of original picture.

| Prog. Type | Sequential | OpenMP | OpenCL |
|---|---|---|---|
| Time (in sec) | 0.51 | 0.05 | 0.96 |

Table 3: Time Elapsed in Image Convolution

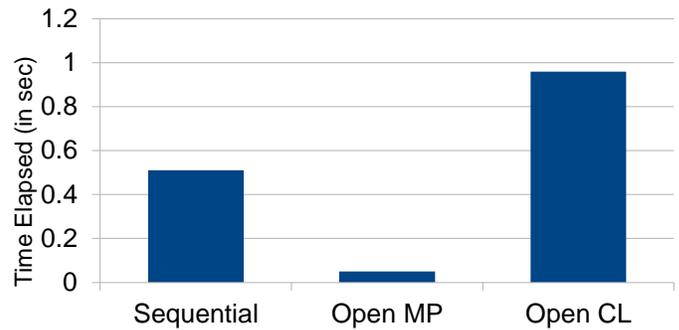

Figure 6: Image Convolution

We took 600 x 400 image, 10 x 10 mask and applied convolution. Table 3 shows time required to process convolution. Performance of OpenMP is better compared to the OpenCL, as evident from the Figure 6.

The speed-up achieved is (Seq/MP) = 0.51/0.05 = 10.2 whereas no speed-up is achieved w.r.t OpenCL as (Seq/CL) = 0.05/0.96 = 0.53.

The convolution algorithm computes the two-dimensional discrete correlation between an image and a template and leaves the result in output image. As a result, OpenMP is much faster compared to OpenCL, as OpenCL is busy in doing background of kernel creation and other things, than actual execution. The actual GPU device execution time can be found by profiling [14] the gpu device which will be very less as compared to OpenMP.

### D. String Reversal

For the comparison purpose, we have taken a string reversal problem. We have considered a huge file in mega-bytes and tried to reverse it using OpenCL.

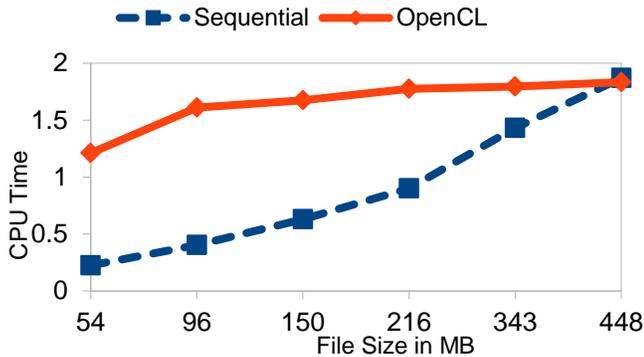

Figure 7 : String Reversal

| File Size | *Sequential* | *OpenCL* |
|---|---|---|
| 54MB | 0.22 | 1.22 |
| 96MB | 0.41 | 1.62 |
| 150MB | 0.63 | 1.68 |
| 216MB | 0.91 | 1.78 |
| 343MB | 1.43 | 1.79 |
| 448MB | 1.87 | 1.84 |

Table 4: String Reversal

String reversal problem is straight forward. Just read the entire file and start copying values from end of the file. As there are no dependencies [15] in this operation, we have not considered OpenMP Programming model. Even if we take OpenMP into consideration, performance will be same as reversing of read string falls in critical section.

From Figure 7 and Table 4, it can be concluded that, OpenCL Programming model is not suitable for this kind of applications.

### VI. RELATED WORK

The present studies dealing with multi-core and GPGPUs have been accelerated by parallelizing matrix-matrix multiplication on a CPU and GPU [16, 17]. In multiple-core clusters systems without GPU accelerators, some contributions have been made to improve the computing power by developing hybrid models [18].

All these studies mentioned above have focused on parallel usage of CPUs and GPUs and have demonstrated significant performance improvement. However, because of the issues related to cache hierarchy, memory and bandwidth in CPU and GPUs, obtaining effective performance evaluation results is an open issue.

### VII. CONCLUSION & FUTURE SCOPE

We studied the behavior of parallel algorithms with respect to OpenMP and OpenCL. The initial results we found were not satisfactory. But, as the number of input data size increased OpenCL gives good performance.

Latest systems are equipped with multi-core architecture. So, OpenMP will be a viable option for cases such as matrix multiplication, image convolution, and other applications. But OpenCL scores well with matrix multiplication.

OpenCL involves a lot of background work like memory allocation, kernel settings and loading, getting platform, device information, computing work-item sizes etc. All this adds overhead in OpenCL. However, we find that, in spite of this overhead, OpenCL gives very good performance. But OpenCL fails in application where it has less scope of work; this can be seen from the string reversal example.

Another finding is that critical section is too expensive. We have implemented OpenMP version of N-Queen problem, but, we find that it has no improvement as only one thread is running at a time. However, we can take advantage of "task" directive in application such as tree traversal.

Overall, we sum up our conclusion as

OpenCL > OpenMP > Sequential

Where > indicates performance. As a future work, we will find algorithms, where OpenMP is more preferable over OpenCL.

Future research work is required in the following problem areas: given an application program, we must check how useful OpenMP or OpenCL is in heterogeneous environment consisting of multiple GPUs and multi-cores. Secondly, a library routine can be developed, which will port application program to CPU using OpenMP or to GPU using OpenCL or combination of these two technologies.

AUTHORS PROFILE

S.R.Sathe received M.Tech in Computer Science from IIT, Bombay (India) and received Ph.D fronm Nagpur University (India). He is currently working as Professor in Department of Computer Science & Engineering at Visvesvaraya National Institute of Technology, Nagpur (India). His research includes parallel processing algorithms, computer architecture.

Krishnahari Thouti, Research Scholar, pursuing Ph.D in Department of Computer Science Engg, at Visvesvaraya National Institute of Techology, Nagpur (India). His area of interest includes parallel processing, compiler and computer architecture.